\documentclass{elsart}

\usepackage{graphicx}
\usepackage{epic,eepic}
\usepackage{amssymb,amsmath}

\begin{document}

\begin{frontmatter}


\title{Method of Monte Carlo grid for data analysis}
\author[wpie,ca]{M.~Filipowicz}
\corauth[ca]{Corresponding author}
\author[jinr]{V.M.~Bystritsky}
\author[swiss]{P.E.~Knowles}
\author[swiss,uiuc]{F.~Mulhauser}
\author[wfitj]{J.~Wo\'zniak}
\thanks[uiuc]{Present address: University of Illinois at
  Urbana--Champaign, USA}

\address[wpie]{Faculty of Fuels and Energy, University of Science and
Technology, al.~Mickiewicza 30, PL--30059 Krakow, Poland}
\address[jinr]{Laboratory of Nuclear Problems, Joint Institute
for Nuclear Research, 141980 Dubna, Moscow Region, Russia}
\address[swiss]{Department of Physics, University of Fribourg, Chemin
du Mus\'ee 3, CH--1700 Fribourg, Switzerland}
\address[wfitj]{Faculty of Physics and Nuclear Techniques, University
of Science and Technology, al.~Mickiewicza 30, PL--30059 Krakow,
Poland}

\begin{abstract}
  This paper presents an analysis procedure for experimental data
  using theoretical functions generated by Monte Carlo.
  Applying the classical chi-square fitting procedure for some
  multiparameter systems is extremely difficult due to a lack of an
  analytical expression for the theoretical functions describing the
  system.
  The proposed algorithm is based on the least square method using a
  grid of Monte Carlo generated functions each corresponding to
  definite values of the minimization parameters.
  It is used for the E742 experiment (TRIUMF, Vancouver, Canada) data
  analysis with the aim to extract muonic atom scattering parameters
  on solid hydrogen.
\end{abstract}

\begin{keyword}
  data fitting \sep Monte Carlo simulation \sep interpolation \sep
  muonic atom \sep scattering \sep E742

  \PACS 02.70.Tt, 02.70Uu, 02.60.Ed, 36.10.Dr, 61.18.-j, 34.50-s
\end{keyword}
\end{frontmatter}

\section{Introduction}

For a wide range of physical problems the only applicable way to
compare experiment with theory is via the Monte-Carlo (MC) method.
Problems of that type are often multiparameter, with nontrivial
interdependencies between the parameters, such as averaging arising
from spatially discrete effects.
Thus, only an exact simulation of the experimental system allows us a
possible analysis, and thus we rely on the MC method.

However, the MC method has several limitations, mostly related to
calculation time and the nature of random sampling.
Provided the simulation has been correctly established to analyze all
competitive processes, the modeling of events which are fairly rare
and hidden inside many other processes often requires long calculation
time to establish sufficient statistics for comparison.
The intrinsic nature of random numbers, and the generators presently
in use, give that any two simulations of the same system can give
different results.
Normally, for large numbers of generated statistics those differences
are small and completely insignificant.
Such limitations are not very important when we make a single
simulation for one experiment, i.e., when there are no variable
parameters and one output result is sufficient.
However, the limitations are amplified when we apply the MC method to
a~fitting procedures.

As is fairly well known, finding the best fit parameters describing a
multitude of data requires repeated calculation of some statistical
estimator.
Most often, the estimator is $\chi^2$, which is defined as the
difference between the theoretical description and the experimental
data, see Eq.~(\ref{eq:single_chisqu}).
The theoretical description of the data depends on several parameters
and thus $\chi^2$ is calculated as a function of those parameters.
Finally, the result is the set of parameters for which the $\chi^2$ is
minimal.

Classical fitting algorithms, e.g., the minimization package
{\sc Minuit}~\cite{james75}, calculate $\chi^2$ from the model
parameters (see Fig.~\ref{fig:fitting_scheme}).
When the $\chi^2$ minimum depends on two or more variables, the error
determination on the parameters as well as the study of the possible
occurrence from several minima require calculations of several
thousands theoretical functions, and hence, the calculations becomes
extremely time-consuming.
However, the MC evaluation of the theoretical function, just for one
set of parameters, is very time exhaustive (measured in hours or even
days):  thousands of iterations are not possible.
Even if the calculation time were acceptable, the intrinsic nature of
MC simulations makes such an approach impossible since instabilities
will arise resulting from the statistical nature of the results.
\footnote{For example, {\sc Minuit} examines whether the theoretical
function is time-independent.
The theoretical function for a~given parameter set is evaluated twice.
If the resulting values are different, the theoretical function is
qualified as a time-dependent and the minimization procedure is
suspended.}.

When fitting, the minimization procedure examines the behaviour of
differences in $\chi^2$ for differing values of the parameter set.
The minimization procedure then calculates the internal gradient of
$\chi^2$ and uses it to control the minimum searching procedure.
The gradient is obtained from a set of partial derivatives for each
variable parameter where the derivatives are calculated numerically
from difference quotients.
Normally, the minimum should be reached when all the gradients
converge to zero.
Clearly, the statistical fluctuations of the MC method can cause
entirely false gradients, and thus such a minimization procedure is
not suitable for our problem.

The work of Zech~\cite{zechx} presents methods for comparing MC
generated histograms to 
experimental data when the analytic distribution is known.
In our case we compare the experimental data with MC
simulations~\cite{wozni96} which including all parameters of the
apparatus, such as the spatial separation of processes, detector
resolution, dead time, etc, and therefore, we can directly compare
experimental and Monte Carlo spectra.

If we ask ``given a data distribution, and a set of MC distributions,
what is the best estimate of the fraction of each MC distribution
present in the data distribution?'' a standard set of
subroutines~\cite{hbook94}, are available to solve that problem.
However, the experimental histograms in our case are not equivalent to
a~summing of discrete MC spectra, and the method above cannot be applied.
An approach~\cite{abbot97} similar to ours was used to determine the
muon energy distributions following muonic capture and atomic cascade
using time of flight methods, although no detailed description of the
method is available.
The aim our paper is to describe such algorithms and to show via
example that it is fully applicable.
As an example the scattering of muonic atoms on a structure of
crystalline hydrogen is presented.

\section{Description of the Method}
\subsection{Modified fitting procedure}

\begin{figure}[ht]
\centerline{\setlength{\unitlength}{1mm}
\thinlines
\tiny
\begin{picture}(120,80)(0,0)
\put(60,75){\oval(25,9)}
\put(56,76){Start,}
\put(52,72){Initialisation}
\put(60,70.5){\vector(0,-1){3}}
\put(27,69){\framebox(20,4)}
\put(28,70){($p_1, p_2, \ldots, p_n$)}
\put(30,67.5){\line(1,0){55}}
\put(2,66){User defined procedure}
\dashline[+30]{3}(5,65)(60,65)
\dashline[+30]{3}(5,65)(5,22)
\dashline[+30]{3}(60,65)(60,22)
\dashline[+30]{3}(5,22)(60,22)
\put(30,67.5){\vector(0,-1){5}}
\put(20,62.5){\line(1,0){25}}
\put(20,62.5){\vector(0,-1){2.5}}
\put(45,62.5){\vector(0,-1){2.5}}
\thicklines
\put(10,40){\framebox(20,20)}
\thinlines
\put(10,50){\line(1,0){20}}
\put(12,58){Using a set of}
\put(12,55){grid functions}
\put(16,52){$\{M_{\gamma,\delta,\ldots}\}$}
\put(12,47){Performing}
\put(12,44){interpolation}
\put(12,41){to find $M$}
\put(35,40){\framebox(20,20)}
\put(37,51){Standard}
\put(37,48){calculation of}
\put(37,45){the function $M$}
\put(20,40){\line(0,-1){5}}
\put(45,40){\line(0,-1){5}}
\put(20,35){\line(1,0){25}}
\put(34,35){\vector(0,-1){3}}
\put(24,25){\framebox(20,7)}
\put(26,28){$\chi^2$ calculation}
\put(34,25){\vector(0,-1){5}}
\put(19,5){\framebox(30,15)}
\put(20,17){Gradient calculation and}
\put(20,13){control of the minimum}
\put(20,9){searching procedure}
\put(49,13){\line(1,0){36}}
\put(85,13){\vector(0,1){10}}
\put(70,23){\framebox(30,30)}
\put(70,44){\line(1,0){30}}
\put(71,50){Checking if the}
\put(71,47){minimum is found}
\put(71,41){Performing all necessary}
\put(71,38){fits and calculations}
\put(71,32){Determining the new set}
\put(71,29){of ($p_1, p_2, \ldots, p_n$)}
\put(85,53){\vector(0,1){10}}
\put(85,63){\line(0,1){4.5}}
\put(100,50){\line(1,0){5}}
\put(105,50){\vector(0,-1){30}}
\put(105,15){\oval(25,10)}
\put(97,15){End of fitting}
\end{picture}}
\caption{Scheme of the fitting procedure using grids. The
        modifications are marked with a thick line. The dashed line
        denotes the procedure defined by the users.}
\label{fig:fitting_scheme}
\end{figure}
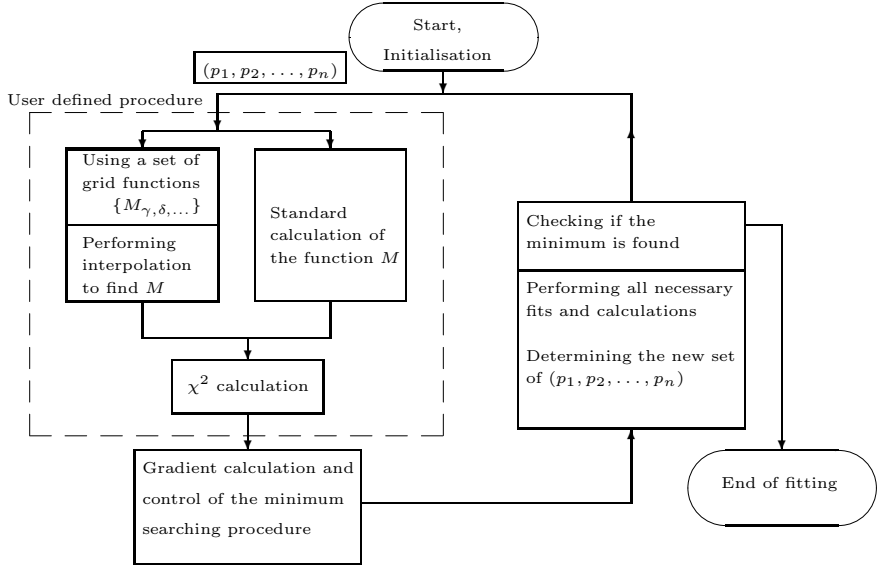

We proposed a modification in the calculation of the theoretical
functions $M(\vec{p})$ which describe the data for a given parameter
 vector $\vec{p}=(p_1,p_2,\ldots,p_n)$.
Before fitting, one generates a set of theoretical functions $\{M\} =
\{M_{\gamma, \delta,\ldots,\phi},$ $M_{\gamma',\delta',\ldots,\phi'},$
$\ldots\}$ for all permutations of a~{\em chosen\/} discrete set of
parameter values $(p_1^{\gamma}, p_2^{\delta}, \ldots, p_n^{\phi} )$
where $\gamma,\delta, \ldots, \phi$ are the function indexes in the
set.
The parameter vector $\vec{p}$ is allowed to assume only discrete
values which gives the grid of theoretical functions a~size $\gamma \times
\delta \times \ldots \times \phi$, and means that the time-consuming
calculations are only executed for a select and limited parameter
set $\{p\}=\{(p_1^{\gamma}, p_2^{\delta}, \ldots,p_n^{\phi}), 
(p_1^{\gamma'}, p_2^{\delta'}, \ldots, p_n^{\phi'}), \ldots \}$.
The resulting set of function values, $\{M\}$, is used to calculate
any theoretical function $M$ for any arbitrary parameter vector
$\vec{p}$ (provided all $p_i$ values in $\vec{p}$ are between some
calculated values of $p_i^{\gamma}$ and $p_i^{\gamma'}$ contained in
the grid) using an interpolation procedure described in
Sec.~\ref{eq:interp_description}.
Once the $\{M\}$ set is known, the interpolation is relatively fast,
and the results, $M(\vec{p})$, can be used to calculate the $\chi^2$.
Note that none of the above precludes $M$ from depending on other
variables, such as time or space, and hence the generated
$M_{\gamma,\delta,\ldots,\phi}$ could just as well be written
$M_{\gamma,\delta,\ldots,\phi}(t,\vec{x})$, so the generated functions
may very well, themselves, be multidimensional.
Figure~\ref{fig:fitting_scheme} presents schematically the fitting
procedure with these modifications.

The number of functions in the set $\{M\}$ depends on each analysis
case and should depend on the behaviour of the function $M(\vec{p})$
for a given parameter $p_i$.
One should note that using too few grid points will give only a~weak
expression of the function's behaviour, whereas using too fine
a~division will, for small parameter changes, falsify the gradient
calculations due to the statistical MC fluctuations.
Properly chosen distances between grid points eliminates the
statistical fluctuations of the theoretical function because the
values between grid points are interpolated.

\subsection{Description of the $\chi^2$ calculation}

We choose MC statistics on average about ten times greater than the
statistical uncertainty in experimental data (less than that and we
are insensitive to our parameters while fitting; more and we use more
MC time for essentially no gain in sensitivity).
Therefore, we neglect the statistical errors connected with the MC and
use the classical $\chi^2$ definition where the fits of the analytical
functions are applied.
Very sophisticated definitions of $\chi^2$, including MC statistical
fluctuations, are presented in Ref.~\cite{kortn03}, however, they are
most useful in the case where experiment and simulation have similar
statistics.

It is possible to use many sets of data from different experimental
conditions, provided they can all be modeled by the same $\vec{p}$
parameters, and define the total $\chi^2$ as the sum of the individual
$\chi^2_k$ calculated separately for a single data set $k$.
Thus, the total $\chi^2$, calculated when we perform simultaneously
fits to $m$ sets of data, is:
\begin{equation}
\chi ^2= m \sum\limits_k {w_k \cdot \chi _k^2 }
\label{eq:multi_chisqu}
\end{equation}
where $m$ is the number of fitted histograms, $k$ the histogram index
running over a single set of data, and $w_{k}$ the corresponding
weight.
The factor $m$ is used to compare the number of degree of freedom
since the chi-squares are non-normalized.

The weights, $w_{k}$, are calculated as a count ratio in each
histogram relative to the total counts in all histograms, such that
histograms with more counts give greater share in the total $\chi^2$:
\begin{equation}
w_k = \frac{\sum\limits_i {N^i (k)} }{\sum\limits_{l=1}^m
           {\sum\limits_i {N^i (l)} } }
\label{eq:weight_k}
\end{equation}
where $N^i$ is the number of events in channel $i$ of the experimental
spectrum.

The partial $\chi_k^2$ is calculated as:
\begin{equation}
\chi_k^2=\sum\limits_i 
\frac{\left[ c_k \cdot M^i(k) - N^i(k) \right]^2}
     {N^i(k)}
\label{eq:single_chisqu}
\end{equation}
where $c_k$ is a factor matching the $k^{\mathrm{th}}$ experimental
$N^i$ with its corresponding MC histograms $M^{i}$ and is given by:
\begin{equation}
c_k=\frac{\sum\limits_i {N^i(k)}}
         {\sum\limits_i {M^i(k)}}
\end{equation}

\subsection{The interpolation method}
\label{eq:interp_description}

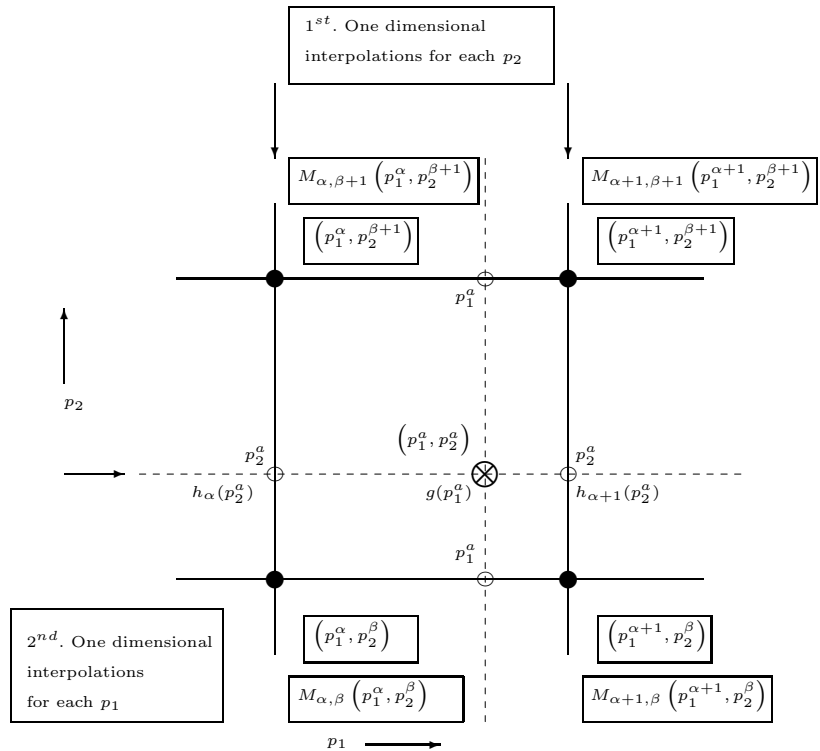
\begin{figure}[hb]
\centerline{\setlength{\unitlength}{1mm}
\thinlines
\tiny
\begin{picture}(120,100)(0,0)
%
\put(40,90){\framebox(35,10)}
\put(42,97){$1^{st}$. One dimensional}
\put(42,93){interpolations for each $p_2$}
\put(38,90){\vector(0,-1){10}}
\put(40,74){\framebox(25,6)}
\put(41,77){$M_{\alpha,\beta+1}\left(p_1^{\alpha}, p_2^{\beta+1}\right)$}
\put(42,66){\framebox(15,6)}
\put(43,69){$\left(p_1^{\alpha}, p_2^{\beta+1}\right)$}
\put(77,90){\vector(0,-1){10}}
\put(79,74){\framebox(31,6)}
\put(80,77){$M_{\alpha+1,\beta+1}\left(p_1^{\alpha+1}, p_2^{\beta+1}\right)$}
\put(81,66){\framebox(18,6)}
\put(82,69){$\left(p_1^{\alpha+1}, p_2^{\beta+1}\right)$}
\put(36.5,62.5){\Large{$\bullet$}}
\put(75.5,62.5){\Large{$\bullet$}}
\put(36.5,22.5){\Large{$\bullet$}}
\put(75.5,22.5){\Large{$\bullet$}}
\put(25,64){\line(1,0){70}}
\put(25,24){\line(1,0){70}}
\put(38,74){\line(0,-1){60}}
\put(77,74){\line(0,-1){60}}
\put(42,13){\framebox(15,6)}
\put(43,16){$\left(p_1^{\alpha}, p_2^{\beta}\right)$}
\put(40,5){\framebox(23,6)}
\put(41,8){$M_{\alpha,\beta}\left(p_1^{\alpha}, p_2^{\beta}\right)$}
\put(81,13){\framebox(15,6)}
\put(82,16){$\left(p_1^{\alpha+1}, p_2^{\beta}\right)$}
\put(79,5){\framebox(25,6)}
\put(80,8){$M_{\alpha+1,\beta}\left(p_1^{\alpha+1}, p_2^{\beta}\right)$}
\put(50,2){\vector(1,0){10}}
\put(45,2){$p_1$}
\put(10,50){\vector(0,1){10}}
\put(10,47){$p_2$}
\put(36.75,37.5){$\bigcirc$}
\put(64.75,63.5){$\bigcirc$}
\put(75.75,37.5){$\bigcirc$}
\put(64.75,23.5){$\bigcirc$}
\put(62,61){$p_1^a$}
\put(62,27){$p_1^a$}
\put(34,40){$p_2^a$}
\put(27,35){$h_{\alpha} (p_2^a)$}
\put(78,40){$p_2^a$}
\put(78,35){$h_{\alpha+1}(p_2^a)$}
\dashline[+30]{1}(20,38)(100,38)
\dashline[+30]{1}(66,80)(66,5)
\put(64,37.5){$\bigotimes$}
\put(54,42){$\left(p_1^a,p_2^a\right)$}
\put(58,35){$g(p_1^a)$}
\put(10,38){\vector(1,0){8}}
\put(3,5){\framebox(28,15)}
\put(5,15){$2^{nd}$. One dimensional}
\put(5,11){interpolations}
\put(5,7){for each $p_1$}
\end{picture}}
\caption{Outlook view of the two-dimensional interpolations. Symbols
        are given in the text.}
\label{eq:interp_scheme}
\end{figure}

The grids are generated only for a finite and discretized set of
parameters for all permutations of the parameters.
However, as follows from the minimization procedure, theoretical
functions are necessary from a continuous parameter space
$(p_1 , p_2, \ldots, p_n)$ and the interpolation
procedure using the grids is applied to generate such functions.
A visual scheme of the two-dimensional interpolation procedure is
presented in Fig.~\ref{eq:interp_scheme}, an example taken from
Ref.~\cite{press92}.
In general, interpolation is only well defined for scalar values.
An interpolation procedure on function can only occurred if we treat
it as a set of scalars.
Therefore the function is given as a table of scalars.
For each table value the interpolation is performed separately.
Then the set of interpolated values give the final interpolated function.

Figure~\ref{eq:interp_scheme} presents the two-dimensional plane for
the parameters $p_1$ and $p_2$, and represents a part of a full grid.
The point at $\left(p_1^a, p_2^a \right)$ for which we wish to find the
function is marked by the sign $\otimes$.
The upper index $a$ means that the variable may takes values from the
continuous spectrum rather than only grid point values.
The four points with index $a$, shown with the open circle symbol
{\footnotesize{$\bigcirc$}}, are intermediate points required by the
calculation. 
Around them are the four grid points,
$\left(p_1^{\alpha},p_2^{\beta}\right)$,
$\left(p_1^{\alpha+1},p_2^{\beta}\right)$,
$\left(p_1^{\alpha},p_2^{\beta+1}\right)$, and
$\left(p_1^{\alpha+1},p_2^{\beta+1}\right)$, denoted with the filled
circle $\bullet$ symbol.
They correspond to the grid functions $M_{\alpha,\beta}$,
$M_{\alpha+1,\beta}$, $M_{\alpha,\beta+1}$, and
$M_{\alpha+1,\beta+1}$, respectively.
The functions $M$ are given in value--channel numerical form, i.e., a
number of counts for each channel of the spectrum.

The idea of a two-dimensional interpolation relies on first performing
$\alpha$ one-dimensional interpolations, where $\alpha$ is the grid
size in this direction, along the directions connecting the grid
points to find the function at point
$\left(p_1^{a},p_2^{\beta}\right)$, for example.
In this way, values for the intermediate function $h_{\alpha}(p_2^a)$
are determined for each node $\alpha$ and $\alpha+1$.
Secondly, a single one-dimensional interpolation along the horizontal
axis (marked with a dashed line) is executed.
To obtain a temporary interpolating function $g(p_1^a)$, one needs a
function $M$ in the point $\left(p_1^a, p_2^a \right)$.
To make the interpolation function $M$ it is necessary to repeat the
described procedure for each tabulated value of the independent
variable\footnote{The central points and widths of the channels have
to be the same for all the grid points.}.
The one-dimensional interpolation is described below.

One uses the following interpolation formula,
\begin{equation}
      h_{\alpha}(p_2^a)=A(p_2^a) \cdot M_{\alpha,\beta} + B(p_2^a) 
      \cdot M_{\alpha,\beta+1} +
      C(p_2^a) \cdot M_{\alpha,\beta}''  
      + D(p_2^{a}) \cdot M_{\alpha,\beta+1}''
\label{eq:interp_formula}
\end{equation}
\begin{equation}
      \begin{split}
        g(p_1^a) & = A(p_1^a) \cdot h_{\alpha}(p_2^a) + B(p_1^a) \cdot
        h_{\alpha+1}(p_2^a) \\
        & + C(p_2^a) \cdot \left[ h_{\alpha}(p_2^a) \right]'' +
        D(p_2^{a})
        \cdot \left[ h_{\alpha+1}(p_2^a) \right]'' \\
      \end{split}
\label{eq:interp_formula_g}
\end{equation}
and finally:
\begin{equation}
      M(p_1^a,p_2^a) = g(p_1^a) \, ,
\label{eq:final_interp}
\end{equation}
where $A$, $B$, $C$, and $D$ are the interpolation coefficients.
The coefficients $A$ and $B$ are defined as
\begin{equation}
      A(x)=\frac{x_{\alpha+1} -x}{\Delta x} \, , 
      \qquad B(x)=\frac{x-x_{\alpha}}{\Delta x}
\label{eq:coeffAB}
\end{equation}
where $x$ is used as a formal notation for the independent variable,
$x \in (x_{\alpha} ,x_{\alpha+1})$.
In our case $x$ plays the role of the parameters $p_1$ or $p_2$.
$\Delta x=x_{\alpha+1} -x_{\alpha}$ are the grid steps.
The coefficients $C$ and $D$ are expressed as:
\begin{equation}
      C(x)=\frac{1}{6}\left( {A^3-A} \right)\cdot \Delta x^2 \, , 
      \qquad D(x)=\frac{1}{6}(B^3-B)\cdot \Delta x^2
\label{eq:coeffCD}
\end{equation}

The dependence of the interpolated function $h$ or $g$ on $x$ in
Eqs.~(\ref{eq:interp_formula}) or~(\ref{eq:interp_formula_g})
is given by a linear dependence on the coefficients $A$, $B$, and a
cubic dependence on the coefficients $C$, $D$.
Thus, this method is called the cubic spline interpolation.

The first step in this method is the calculation and tabulation of the
second derivative values for all functions $\{M_{\alpha,\beta}''\}$ in
the grid.
If one wants to use Eq.~(\ref{eq:interp_formula_g}) the second
derivative of ${h}''$ is required.
One obtains the second derivatives by solving the following expressions
\begin{equation}
      M_{\alpha-1,\beta}'' + 4 M_{\alpha,\beta}'' 
      + M_{\alpha+1,\beta}'' = 6 \cdot
      \frac{M_{\alpha+1,\beta}+M_{\alpha-1,\beta}}{\Delta x^2} \, ,
\label{eq:second_deriv}
\end{equation}
This expression is correct only when grid points are spaced equally.
Based on Eq.~(\ref{eq:second_deriv}), a triangular matrix is created
and reduced by using a suitable numerical algorithm.
To solve Eq.~(\ref{eq:second_deriv}) one needs boundary conditions.
In the presented analysis, the second derivatives for the first and
last values of $M''$ are set to zero, the so-called natural cubic
spline interpolation.

\begin{figure}[!htbp]
\centerline{\includegraphics[width=0.9\linewidth]{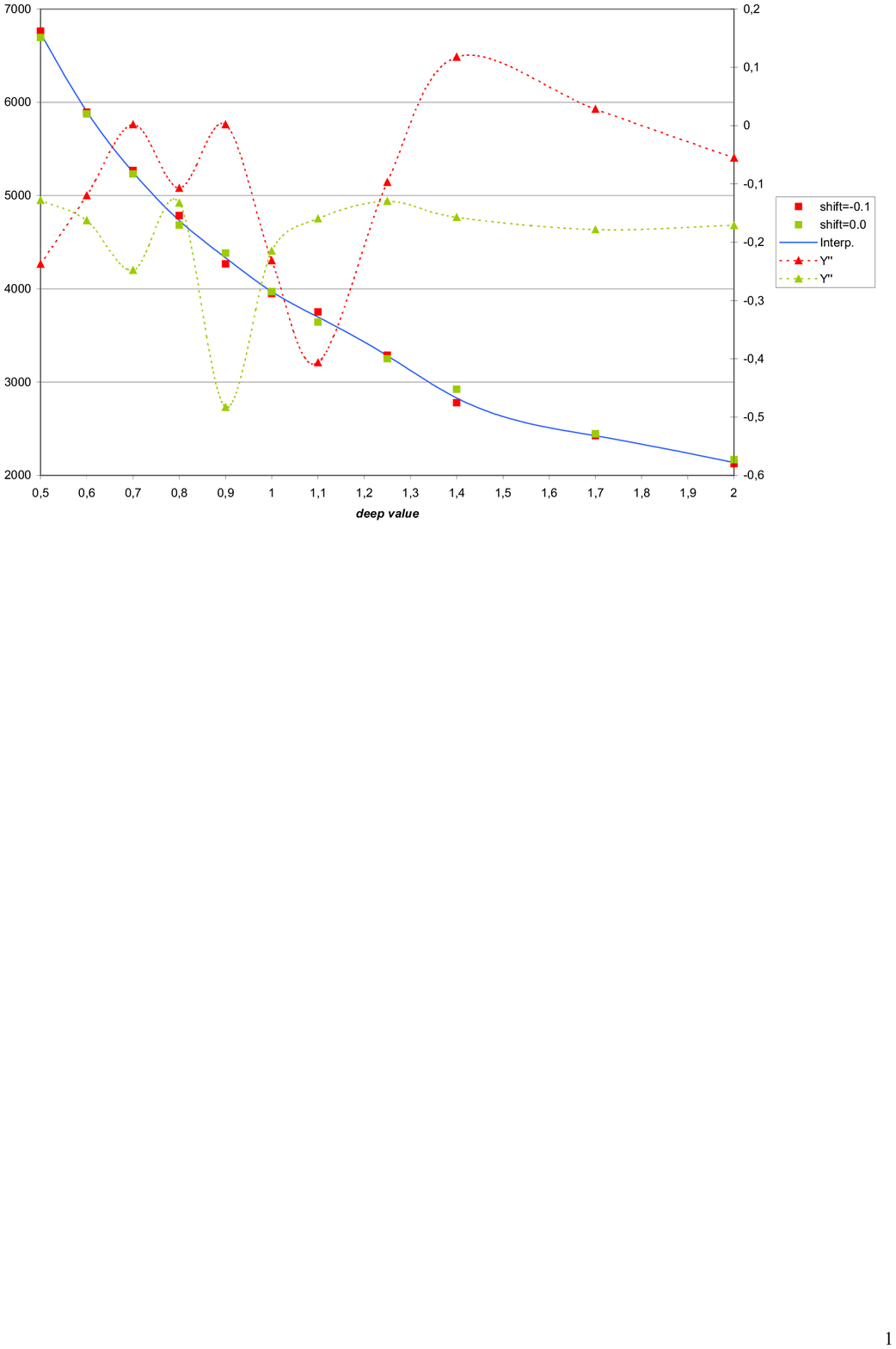}}
\caption{Graph of two functions, $Y(\mbox{shift}=-0.1\,\mbox{eV},
         \mbox{depth})$ given by the red squares and
         $Y(\mbox{shift}=-0.0\,\mbox{eV}, \mbox{depth})$ with the
         green squares.  The red triangles show
         $Y''(\mbox{shift}=-0.1\,\mbox{eV}, \mbox{depth})$ , whereas
         the green ones stand for $Y''(\mbox{shift}=0.0\,\mbox{eV},
         \mbox{depth})$. The green solid line is obtained by an
         interpolation function for a shift value of $-0.05$~eV\@.}
\label{fig:example_method}
\end{figure}

The ability of our interpolation routine to reproduce a function at
$(\alpha,\beta)$ using the four points $(\alpha\pm1,\beta\pm1)$ is
shown in Fig.~\ref{fig:example_method} for a case where the two 
functions $Y(\mbox{shift}, \mbox{depth})$ are given.

\section{Application of the method}
\subsection{Description of the experiment}

\begin{figure}[!htbp]
\centerline{\includegraphics[width=0.9\linewidth]{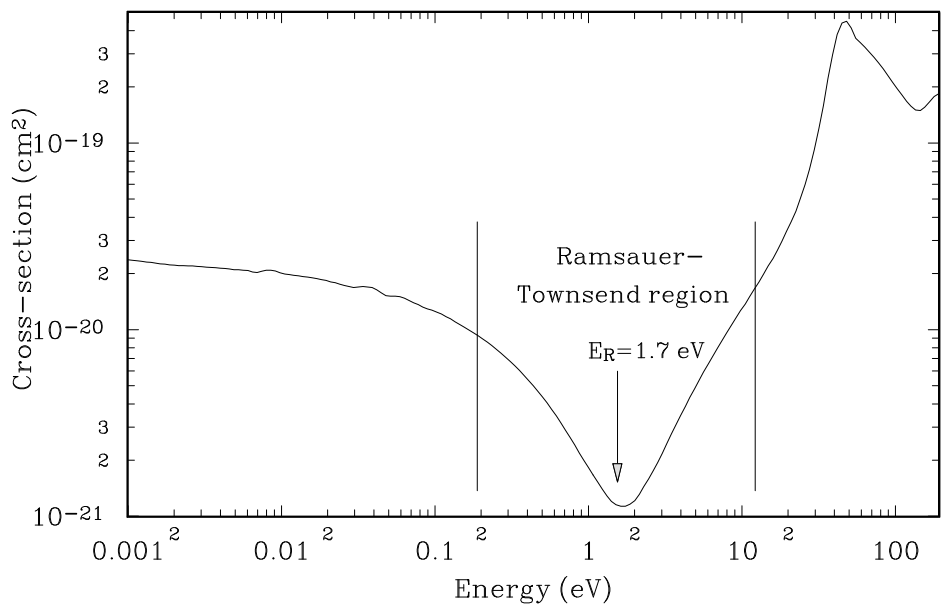}}
\caption{Elastic scattering cross-sections for $d\mu+p$ and R--T
        region. The essential values used in the parametrization are
        given.}
\label{fig:cross_section}
\end{figure}

In this section we apply our analysis method to the data obtained in
the E742 experiment performed at TRIUMF, Vancouver (Canada).
The experiment was dedicated to the study of $\mu$-atomic processes
occurring in solid hydrogen isotopes.
It is of particular interest to obtain the characteristics of the
interacting systems (collision energy of muonic atoms with a
crystalline structure).
We want to reconstruct the energy dependence of the elastic scattering
cross-sections for muonic atoms in the process: $d\mu+p \to d\mu+p$ on
crystalline hydrogen at a temperature of 3~K\@.
Theoretical calculations~\cite{chicc92} postulated that there exists
an energy region of abnormally small cross-section called
the Ramsauer--Townsend (R--T) region.
Figure~\ref{fig:cross_section} shows this dependence and the R--T region
is visible.
The aim of the measurement was to find experimentally the R--T region
and verify the theoretical prediction.
The accuracy of the theoretical calculations was low for energies inside
the R--T region and the R--T minimum was also determined by theoretical
calculations of poor accuracy.
We assumed only that the general shape of the cross-section curve was valid.
We vary only the depth and the position of the minimum of the R--T
region.

The most accurate experimental method would be the use of
a~(selectable) monoenergetic beam of muonic $d\mu$ atoms, and, by
aiming the beam at a~thin foil of crystalline hydrogen and (like in
the Rutherford experiment) detecting the intensity and energy of the
scattered $d\mu$ atoms would allow us to determine the cross-section
as a function of the $d\mu$-atoms energy.
The method would also give us the scattering angles, and interpreting
the data would be easy.
However, even in this maximally simplified case it is most probable
that the MC method have to be used.

Unfortunately, it is impossible to use the direct method because such a
source of $d\mu$ atoms does not exist and there are no detectors
measuring directly the energy of muonic $d\mu$ atoms.
Therefore, in the real experiment we produced muonic atoms inside a
structure of solid hydrogen, starting a~time counter on the muon
arrival.
The energy of the scattered $d\mu$ atoms after leaving the crystal is
measured indirectly via a time of flight method.
The muonic atom flew between two layers placed at some distance
between each other in vacuum.
The first layer is a~source of muonic atoms and is
treated as an {\em emitter} of $d\mu$-atoms.
The second layer is covered by neon and is treated as the {\em
detector} of muonic atoms: a~$d\mu$ atom entering the neon layer
transfers the muon to the Ne almost immediately yielding an x~ray 
which determined the stop of the time counter.
Detailed information about the experiment can be found in
Ref.~\cite{mulha01} and references therein.

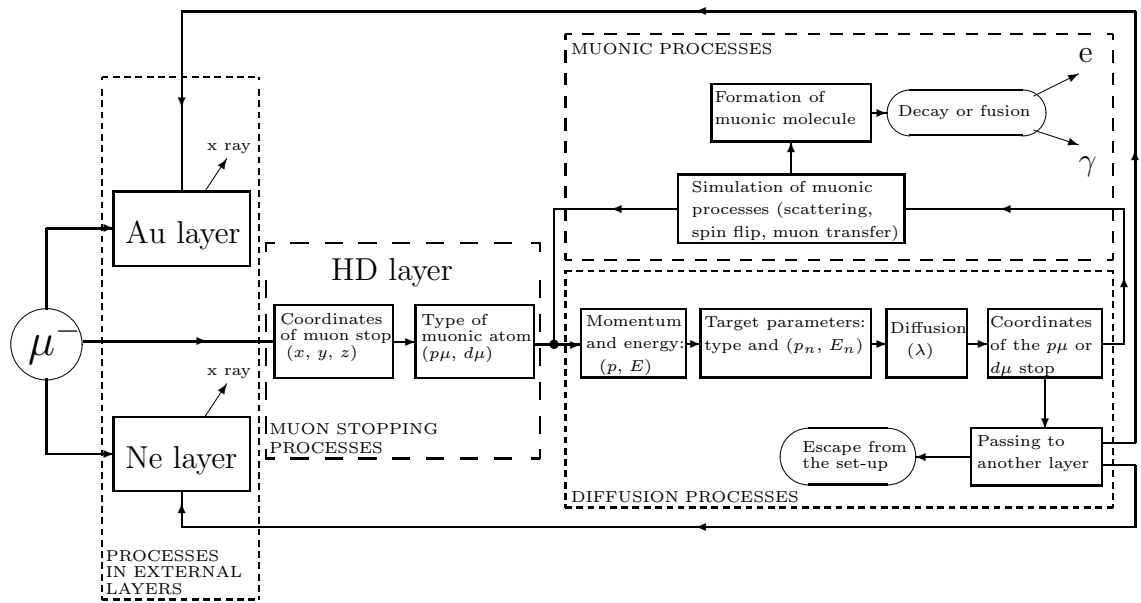
\begin{figure}[htbp]
\centerline{
\setlength{\unitlength}{0.75mm}
\linethickness{0.4pt}
\ifx\plotpoint\undefined\newsavebox{\plotpoint}\fi 
\begin{picture}(205,110)(0,0)
\put(8,46){\circle{12.75}}
\put(4.8,44){\Large $\mu^-$}
\put(8,52.2){\line(0,1){14.3}}
\put(8,66.5){\vector(1,0){12}}
\put(8,39.8){\line(0,-1){13.3}}
\put(8,26.5){\vector(1,0){12}}
\put(17.87,0.69){\dashbox{1}(27.75,92.71)[cc]{}}
\put(18.5,7.63){\tiny PROCESSES}
\put(18.5,4.63){\tiny IN EXTERNAL}
\put(18.5,1.63){\tiny LAYERS}
\put(20,60){\framebox(24,13)[cc]{}}
\put(22,64){Au layer}
\put(36,73){\vector(2,3){4}}
\put(36.5,80.5){\tiny x~ray}
\put(20,20){\framebox(24,13)[cc]{}}
\put(22,24){Ne layer}
\put(36,33){\vector(2,3){4}}
\put(36.5,40.5){\tiny x~ray}
\put(58.44,57.45){HD layer}
\put(46.88,25.71){\dashbox{3}(48.56,38.26)[cc]{}}
\put(47.51,30){\tiny MUON STOPPING}
\put(47.51,26.71){\tiny PROCESSES}
\put(14.5,46.5){\line(1,0){34.0}}
\put(34.5,46.5){\vector(1,0){2.0}}
\put(48.5,40){\framebox(20.8,12.8)[cc]{}}
\put(49.5,49.67){\tiny Coordinates}
\put(49.5,46.67){\tiny of muon stop}
\put(50.5,43.67){\tiny ($x$, $y$, $z$)}
\put(69.3,46.4){\vector(1,0){4.2}}
\put(73.5,40){\framebox(20.8,12.8)[cc]{}}
\put(74.5,49.52){\tiny Type of}
\put(74.5,46.52){\tiny muonic atom}
\put(74.5,43.52){\tiny ($p\mu$, $d\mu$)}
\put(94.3,45.9){\vector(1,0){8.4}}
\put(100,17){\dashbox{1}(97,42)[cc]{}}
\put(101,18){\tiny DIFFUSION PROCESSES}
\put(102.8,40.2){\framebox(18.5,12)[cc]{}}
\put(103.65,49.3){\tiny Momentum}
\put(103.65,45.3){\tiny and energy:}
\put(106,41.3){\tiny ($p$, $E$)}
\put(121.3,46){\vector(1,0){2.7}}
\put(124,40.2){\framebox(30,12)[cc]{}}
\put(124.6,49){\tiny Target parameters:}
\put(124.6,45){\tiny type and ($p_n$, $E_n$)}
\put(154,46){\vector(1,0){3}}
\put(157,40.2){\framebox(14,12)[cc]{}}
\put(157.6,48){\tiny Diffusion}
\put(160.5,44){\tiny ($\lambda$)}
\put(171,46){\vector(1,0){4}}
\put(175,40.2){\framebox(20,12)[cc]{}}
\put(175.33,49){\tiny Coordinates }
\put(175.33,45){\tiny of the $p\mu$ or}
\put(175.33,41){\tiny $d\mu$ stop}
\put(185,40.2){\vector(0,-1){9.2}}
\put(172,21){\framebox(23,10)[cc]{}}
\put(173,28){\tiny Passing to}
\put(173,24){\tiny another layer}
\put(172,26){\vector(-1,0){10}}
\put(150,26){\oval(24,10)[]}
\put(142,27){\tiny Escape from}
\put(142,24){\tiny the set-up}
\put(195,24.5){\line(1,0){6}}
\put(201,24.5){\line(0,-1){11}}
\put(201,13.5){\line(-1,0){169}}
\put(125,13.5){\vector(-1,0){2}}
\put(32,13.5){\line(0,1){6.5}}
\put(32,16){\vector(0,1){2}}
\put(195,28.5){\line(1,0){6}}
\put(201,28.5){\line(0,1){76.5}}
\put(201,78.5){\vector(0,1){2}}
\put(201,105){\line(-1,0){169}}
\put(125,105){\vector(-1,0){2}}
\put(32,105){\line(0,-1){32}}
\put(32,90){\vector(0,-1){2}}
\put(195,46){\line(1,0){4}}
\put(199,46){\line(0,1){24}}
\put(199,56){\vector(0,1){2}}
\put(199,70){\line(-1,0){39}}
\put(179,70){\vector(-1,0){2}}
\put(100,61){\dashbox{2}(97,40)[cc]{}}
\put(101,97.5){\tiny MUONIC PROCESSES}
\put(120,64){\framebox(40,12)[cc]{}}
\put(122.4,73){\tiny Simulation of muonic}
\put(122.4, 69){\tiny processes (scattering,}
\put(122.4, 65){\tiny spin flip, muon transfer)}
\put(140,76){\vector(0,1){6}}
\put(120,70){\line(-1,0){22}}
\put(110,70){\vector(-1,0){2}}
\put(98,70){\line(0,-1){24}}
\put(126,82){\framebox(28,10)[cc]{}}
\put(126.5,89){\tiny Formation of}
\put(126.5,85){\tiny muonic molecule}
\put(154,87){\vector(1,0){3}}
\put(171,87){\oval(28,8)[]}
\put(159,86.5){\tiny Decay or fusion}
\put(183,84){\vector(3,-1){8}}
\put(183,90){\vector(2,1){8}}
\put(191,96){e}
\put(191,77){$\gamma$}
\put(98,46){\circle*{1.5}}
\end{picture}}
\caption{Scheme of FOW program used to simulate the E742
        experiment. Four process blocks are shown, namely the
        processes in external layers, the muon stopping, the
        diffusion, and the muonic processes.}
\label{fig:fow_programme}
\end{figure}

To analyze the experiment, we have to use MC simulations because other
processes are competing with the scattering, as well as the
complications coming from the geometry of the experiment.
Figure~\ref{fig:fow_programme} presents the scheme of the processes taken
into account~\cite{wozni96}.
The initial time is given when a muon enters the apparatus.
The output of the simulation is a x--ray time spectrum (see processes
in external layers) which can be directly compared with the
experimentally measured one.

\subsection{The grid construction}

The grid method is illustrated for one chosen set of three experiments
with different experimental conditions.
Since the scattering cross-section does not depend on experimental
conditions, fitting three different conditions with the same
parameters should reduce systematics.
The theoretical dependence of the cross-section from the collision
energy was parametrized on two ways.
One changes the position of the minimum on the energy axis (varying
$\Delta E$) and the depth of R--T minimum (varying $s$).
In our example $p_1$ represents the shift of R--T energy ($\Delta E$)
as can be seen in Figs.~\ref{fig:cross_section}
and~\ref{fig:shift_parametrisation}.
The parameter $p_2$ represents a rescaling factor $s$ of the minimal
cross-section value (see Figs.~\ref{fig:cross_section}
and~\ref{fig:deep_param}).
The function $M$ is the MC~time spectrum of muonic atoms reaching the
neon layer for $\Delta E$ and $s$.
Thus, the grid is defined as $\{M_{\Delta E_\alpha, s_\beta}\}$.

\subsubsection{Change of the R--T minimum energy}

\begin{figure}[ht]
\centerline{\includegraphics[width=0.9\linewidth]{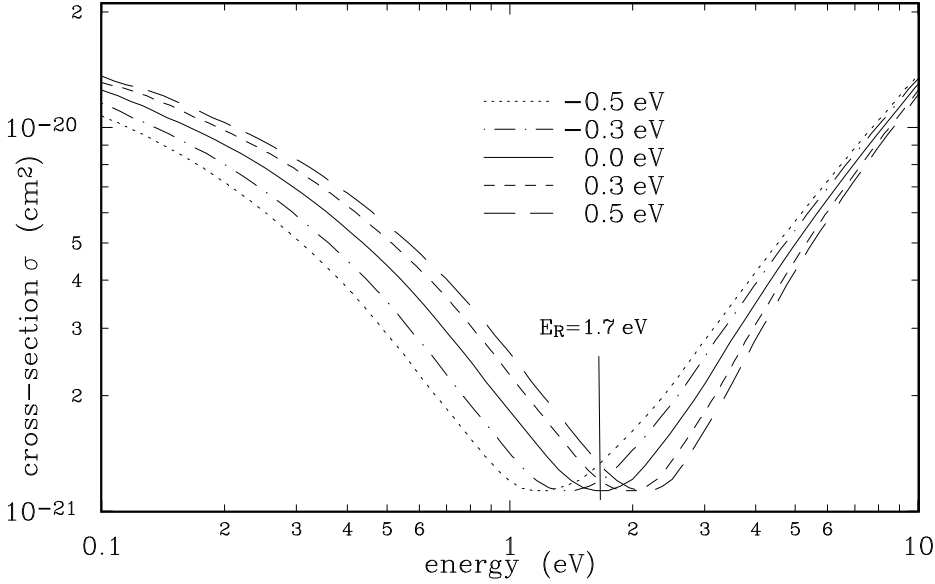}}
\caption{$\Delta E$-parametrization of the R--T position minimum.
        Curves for some value of $\Delta E$ are shown.}
\label{fig:shift_parametrisation}
\end{figure}

\begin{table}[b]
\caption{Characteristic values used in the cross section
        parametrization.}
\label{tab:energies}
\begin{center}
\begin{tabular}{llll} 
Notation  & Name & Value & Units \\ 
$E_{R }$  & Energy of R--T minimum & 1.7 & eV \\ 
$E_{min}$ & Minimum of energy range & 0.001 & eV \\ 
$E_{max}$ & Maximum of energy range & 190.5 & eV \\ 
$\sigma(E_R)$ & Minimum cross section for the R--T energy & $1.13
\times 10^{-21}$ & $\mathrm{cm}^2$ \\ 
\end{tabular}
\end{center}
\end{table}

The energy axis is transformed according to $E \to E_{tr}$ and then
$\sigma(E) \to \sigma(E_{tr})$.
The energy transformation are done via
\begin{equation}
E_{tr} =\left\{ {{
        \begin{array}{*{20}c}
        {\begin{array}{l}
          E+ \frac{E-E_{\min } }{E_R -E_{\min } }\cdot \Delta
          E,\qquad \mbox{for} \quad E \le E_R \\ \\
         \end{array}} \hfill \\
        { E+ \frac{E_{\max } -E}{E_{\max } -E_R }\cdot \Delta 
          E, \qquad \mbox{for} \quad E > E_R } \hfill \\
        \end{array} }} \right.
\label{eq:shift_change}
\end{equation}
where $\Delta E$ is the energy shift, $E$~the unshifted energy, and
$E_{R}$ the original value of the R--T energy minimum.
Limits $E_{min}$ and $E_{max}$ define the range where the
transformation is applied.
For $E=E_R$ the transformed energy $E_{tr}$ given by
Eq.~(\ref{eq:shift_change}) becomes $E_{tr}= E+ \Delta E$, and thus,
this parametrization can be treated as a shift.
Characteristic values for theses variables are given in
Table~\ref{tab:energies}.

\begin{figure}[ht]
\centerline{\includegraphics[width=0.9\linewidth]{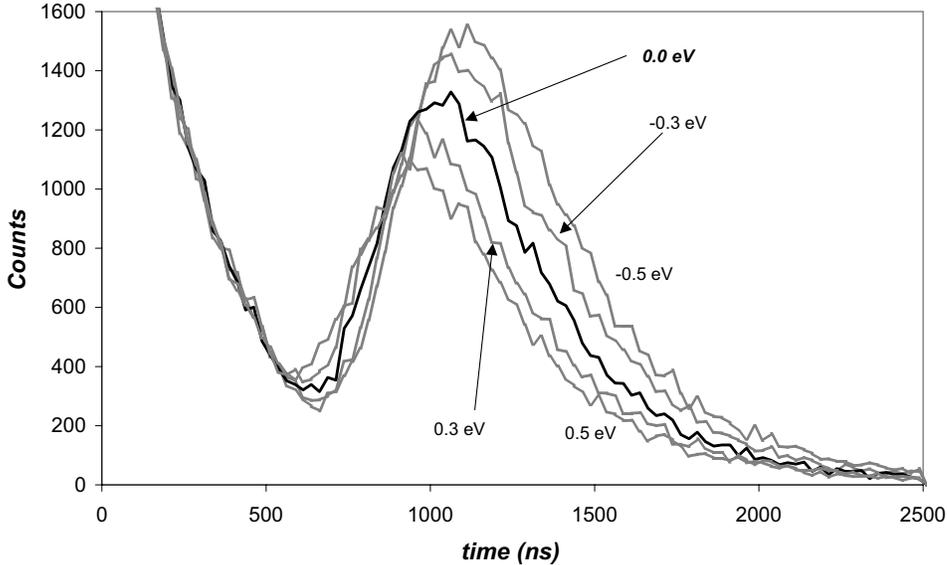}}
\caption{$M_{\alpha}(\Delta E)$, the time spectra for the
        shift-parametrization and $s=1$. Curves are shown for the same
        values as given in Fig.~\ref{fig:shift_parametrisation}.}
\label{fig:MC_deuter_shift}
\end{figure}

The shift values, $\Delta E$, were defined as 11 points between
$-0.5$~eV to $0.5$~eV, in steps of $0.1$~eV\@.
These values were chosen as a result of previous tests of the
experimental data with different shift values.
The shape of the cross-section curves are presented on
Fig.~\ref{fig:shift_parametrisation}.
The resulting MC time spectra for such parametrized cross-section are
presented in Fig.~\ref{fig:MC_deuter_shift}.

\subsubsection{Change of the R--T depth minimum}

\begin{figure}[!htbp]
\centerline{\includegraphics[width=0.9\linewidth]{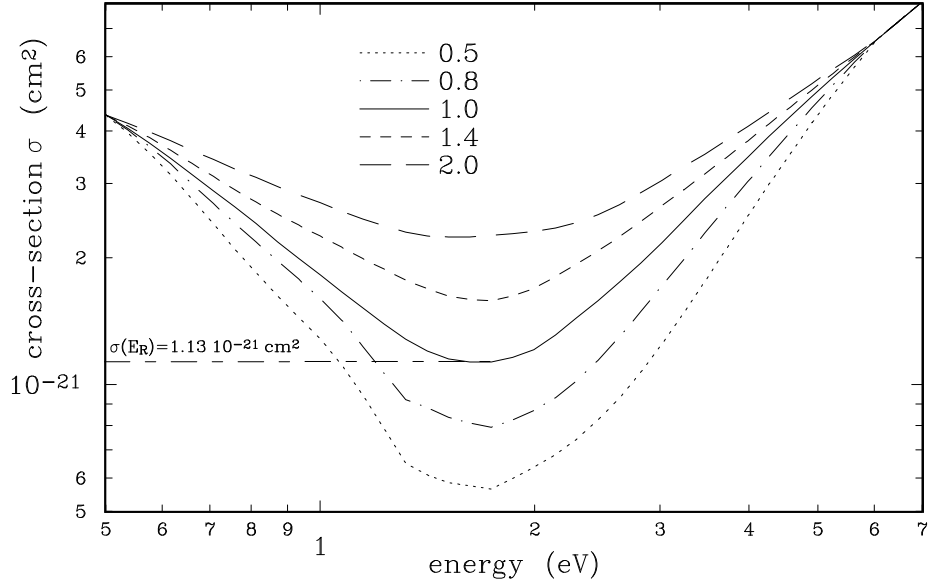}}
\caption{Depth-parametrization of the R--T minimum depth. Some values
        of $s$, which concern the minimum cross section $\sigma(E_R)$
        are indicated.}
\label{fig:deep_param}
\end{figure}

The curves for the depth parametrization have a rescaled minimum
cross-section $\sigma(E_R)= s \times \sigma(E_R)$.
The rescaling took place in the energy range $0.5-6$~eV\@, with 11
values of the rescaling parameter $s$ chosen from 0.5 to 1.1 by steps
of 0.1, with additional values of 1.25, 1.4, 1.7, and 2.0, as can be
seen in Fig.~\ref{fig:deep_param}.

\begin{figure}[!htbp]
\centerline{\includegraphics[width=0.9\linewidth]{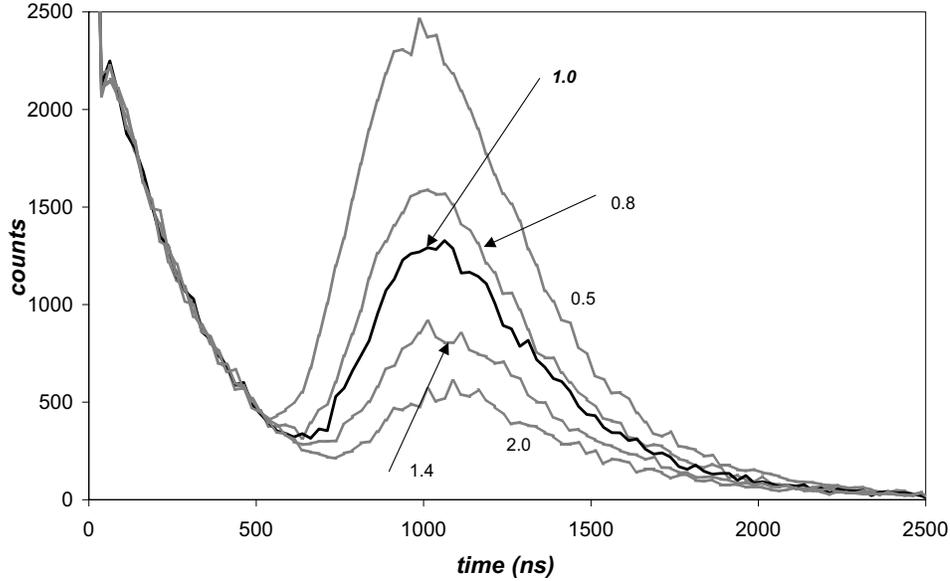}}
\caption{$M_{\beta}(s)$, the time spectra for the depth
        parametrization, without the energy shift $\Delta E=0$.}
\label{fig:MC_deuter_depth}
\end{figure}

To preserve the smooth form of the cross sections, values within the
0.5--6~eV range were not globally scaled by~$s$, but were scaled by a
factor which ranged from~1 at the borders, to $s$ at the energy of the
R--T minimum.
The $s$ values (except the minimum value) were selected numerically to
reproduce the characteristic shape of the cross-section function
inside the R--T region, as shown in Fig.~\ref{fig:MC_deuter_depth}.
The MC time spectra for the depth parametrized cross-sections are
presented in Fig.~\ref{fig:MC_deuter_depth}.
The functions $M_{\alpha}(\Delta E)$ and $M_{\alpha}(s)$ given in
Figs.~\ref{fig:MC_deuter_shift} and~\ref{fig:MC_deuter_depth},
respectively, were combined and a grid $\{M_{\alpha,\beta}(\Delta
E,s)\}$ was created.

\section{The results}

\begin{figure}[hb]
\centerline{\includegraphics[width=0.9\linewidth]{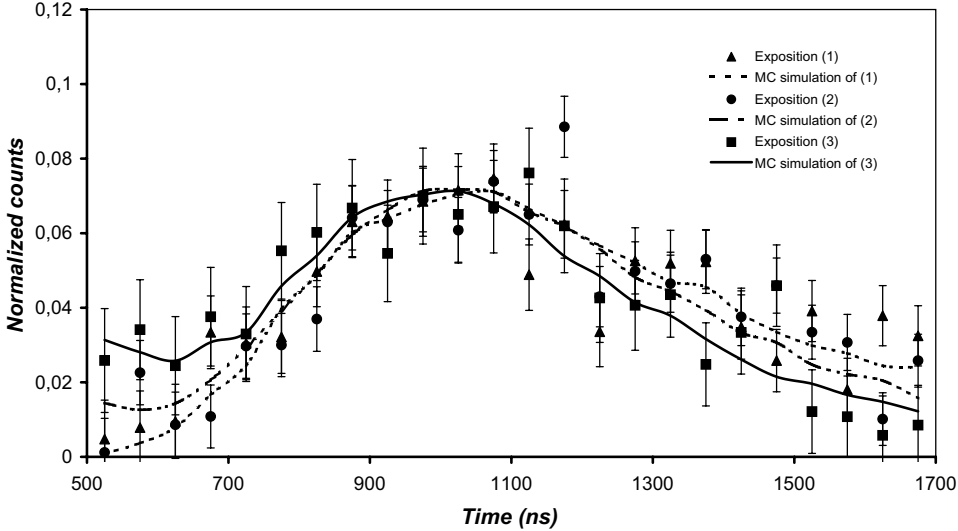}}
\caption{Example of a fit.}
\label{fig:example_fit}
\end{figure}

An example of the experimental data fits using the MC time spectra
(see Figs.~\ref{fig:MC_deuter_shift} and \ref{fig:MC_deuter_depth}) is
presented in Fig.~\ref{fig:example_fit}.
In this case three sets of experimental data (called Expositions 1--3)
were fitted.
Fits were performed for a number of data combinations.

The average values from all possible fit combinations give the final
result:
\begin{equation}
    s = 1.12 \pm 0.20 \, , \qquad \Delta E = 0.30 \pm 0.14 \, \mbox{eV.}
\end{equation}
The errors are connected with low experimental statistics, some
background problems, and also with the grid steps (0.10~eV for $\Delta
E$ and $0.1-0.15$ for $s$) and finally with the interpolation
procedure.
The result means that our experimental result confirms the theoretical
cross-section energy dependence in the R--T region, but indicates that
the minimum of the cross-section value occurs for energies higher than
predicted, at about 2~eV instead of 1.7~eV in
Fig.~\ref{fig:cross_section}.
The absolute value of the fitted cross-sections agree with the
theoretical value.

\section{Conclusion}

The method allowed us to perform a correct comparison of experimental
data with theoretical predictions based on MC calculations.
Although the experimental data was obtained in only a few weeks of
muon beam usage,  the grid construction was a time-consuming step
requiring more than six months of calculation.
The fitting procedure was quick and allowed us to prepare many fits
for any combination of the data and perform more complex analysis of
the data themselves, e.g., $\chi^2$--contour and error calculations.

To establish a more precise set of mathematical rules which test the
correctness of this method, one needs to perform further studies but
such was not the aim of this work.
The grid method, as demonstrated here, is fully acceptable for
analyzing systems with complex multiparameter dependences.
Our procedure was internally checked by comparing results of fitting
single and summed data.
The errors of single data fits are bigger but they lie within the
range of the experimental errors of summed data.

\section*{Acknowledgments}

The authors thank ACK CYFRONET in Krakow for allowing them to use
their supercomputers.
%
This work was supported by the Russian Foundation for Basic Research,
Grant No.~01--02--16483, the Polish State Committee for Scientific
Research, the Swiss National Science Foundation, TRIUMF, and a grant
WPiE/AGH No.~10.10.210.52/7.

\bibliographystyle{elsart-num}
\bibliography{mucf}

\end{document}